\documentclass[conference]{IEEEtran}
\IEEEoverridecommandlockouts
\usepackage{cite}
\usepackage{amsmath,amssymb,amsfonts}
\usepackage{algorithmic}
\usepackage{graphicx}
\usepackage{textcomp}
\usepackage{xcolor}
\usepackage{subfigure}
\usepackage{float}
\usepackage{subfig}
\usepackage{keyval}
\usepackage{array,makecell}
\usepackage{wrapfig}
\usepackage{booktabs}

\def\BibTeX{{\rm B\kern-.05em{\sc i\kern-.025em b}\kern-.08em
    T\kern-.1667em\lower.7ex\hbox{E}\kern-.125emX}}
\begin{document}

\title{EM-TTS: Efficiently Trained Low-Resource Mongolian Lightweight Text-to-Speech\\
}

\author{\IEEEauthorblockN{Ziqi Liang$^{1 \dagger}$, Haoxiang Shi$^{1 \dagger}$\thanks{$\dagger$ Equal Contributions}, Jiawei Wang$^{2}$, Keda Lu$^{1\ast}$\thanks{$^\ast$ Corresponding author: Keda Lu (lukeda@mail.ustc.edu.cn)}}
\IEEEauthorblockA{\textit{$^{1}$University of Science and Technology of China}\\\textit{$^{2}$Hefei University of Technology}}
}


\maketitle

\begin{abstract}
Recently, deep learning-based Text-to-Speech (TTS) systems have achieved high-quality speech synthesis results. Recurrent neural networks have become a standard modeling technique for sequential data in TTS systems and are widely used. However, training a TTS model which includes RNN components requires powerful GPU performance and takes a long time. In contrast, CNN-based sequence synthesis techniques can significantly reduce the parameters and training time of a TTS model while guaranteeing a certain performance due to their high parallelism, which alleviate these economic costs of training.
In this paper, we propose a lightweight TTS system based on deep convolutional neural networks, which is a two-stage training end-to-end TTS model and does not employ any recurrent units.
Our model consists of two stages: Text2Spectrum and SSRN. The former is used to encode phonemes into a coarse mel spectrogram and the latter is used to synthesize the complete spectrum from the coarse mel spectrogram. Meanwhile, we improve the robustness of our model by a series of data augmentations, such as noise suppression, time warping, frequency masking and time masking, for solving the low resource mongolian problem.
Experiments show that our model can reduce the training time and parameters while ensuring the quality and naturalness of the synthesized speech compared to using mainstream TTS models. Our method uses NCMMSC2022-MTTSC Challenge dataset for validation, which significantly reduces training time while maintaining a certain accuracy.
\end{abstract}

\begin{IEEEkeywords}
Text-to-Speech, Sequence-to-Sequence, Efficiently, Lightweight.
\end{IEEEkeywords}

\section{Introduction}
At present, the speech synthesis technology of mainstream languages such as Chinese and English has developed relatively maturely, and the speech synthesis of low-resource languages has gradually attracted more and more attention. Mongolian is the most famous and widely spoken language among the Mongolian people. Worldwide, there are about 6 million users. At the same time, Mongolian is also the main national language of China's Inner Mongolia Autonomous Region \cite{ref_article1}. Therefore, the study of Mongolian-oriented speech synthesis technology is of great significance to the fields of education, transportation, and communication in minority areas.

Traditional speech synthesis methods mainly include speech synthesis techniques based on waveform splicing and statistical parametric acoustic modeling. 
In recent years, deep neural network-based Text-to-speech models, such as those using RNN structures \cite{ref_article5,ref_article6} have achieved high-quality results. People gradually began to study to reduce their dependence on manual features and models started relying solely on mel or linear spectrograms. It does not use other speech parameters such as fundamental frequency, formant parameters, etc., but only focuses on the frequency spectrum to represent the audio signal, among which Tacotron \cite{ref_article7} is more typical. 
The Tacotron series of models are decoded with the output of the previous moment as the input of the next moment for acoustic parameter prediction. 
However, using this autoregressive decoding structure model as an acoustic model has a drawback in that it uses many recursive components with high training costs, which is demanding on GPU computing resources. 

In order to improve the decoding speed, researchers further propose speech synthesis models based on non-autoregressive acoustic modeling such as FastSpeech series \cite{fastspeech1, ref_article10, ref_article11}. These non-autoregressive acoustic models can take a given text as input and output the whole sequence of acoustic parameters in parallel, without relying on the acoustic parameters obtained from the decoding of historical moments. However, non-autoregressive models based on transformer often lack the ability to model local features and extract sequence features step by step, and have slow convergence speeds.
In contrast, CNN can effectively capture local features and reduce the correlation between model parameters, thus accelerating the convergence speed of gradient descent. In addition, higher-level speech features can also be extracted step by step through the stacking of convolutional layers, allowing the model to better understand the latent speech features.


In this paper, we propose the Mongolian Text-to-Speech model EM-TTS, which is a sequence2sequence model based entirely on CNN modules. Our contributions of this paper are summarized as follows:
\begin{itemize}
\item We design an lightweight acoustic model including two stages of Text2Spectrum and SSRN. The former is used to encode phonemes into coarse mel spectrogram, the latter is used to synthesize the complete spectrum from the coarse mel spectrogram. This design ensures synthesis quality while significantly reducing model parameters and training time.
\item We introduce data augmentation means such as time mask and frequency mask to obtain new audio samples, which increases the training samples to deal with the current situation of low-resource Mongolian language data.
\end{itemize}

\section{Related Work}

Recently, deep learning-based Text-to-Speech has been extensively researched, yielding surprisingly high-quality results in some recent studies. The TTS process mainly includes two parts, one is the conversion from text transcription to spectrogram using the acoustic model, and the other is the conversion from spectrogram to speech waveform using the vocoder, such as \cite{ref_article10, fastspeech1, parallelGAN}.

Afterward, some parallel end-to-end methods achieved high-quality speech synthesis, reducing the information loss in text-to-mel spectrum conversion such as VITS \cite{vits}, YourTTS \cite{yourtts}, NaturalSpeech \cite{naturalspeech}.
Later, people introduce language modeling into TTS and train a neural codec language model using discrete codes derived from off-the-shelf neural audio codec models. some works \cite{valle, speartts} treat TTS as a conditional language modeling task rather than regressing continuous signals as in previous work. With the application of neural codecs, a lot of works cast TTS as a composition of two sequence-to-sequence tasks: from text to high-level semantic tokens and from semantic tokens to low-level acoustic tokens.

In addition, with the application of diffusion models, diffusion-based TTS work has also emerged and demonstrated high-quality synthesis effects, such as \cite{gradtts, difftts, guidetts, PriorGrad}. Diffusion-based TTS models usually use diffusion as a postnet for feature enhancement. Taking Grad-TTS \cite{gradtts} as an example, it includes a feature generator and a score-based decoder using diffusion probabilistic modeling, which reconstructs the Gaussian noise output into a high-quality mel spectrogram by gradually transforming noise predicted by the encoder and aligning with text input by means of Monotonic Alignment Search.

However, training a high-quality TTS is particularly difficult in the face of low-resource data. \cite{zhang2022semi, zhang2022tdass} have explored semi-supervised learning and domain adaptation methods, but there are few methods to solve low-resource problems from the perspective of training data.
For vocoder research, a neural network-based vocoder has also been proposed to directly model the speech sample points. The neural vocoder directly learns the mapping relationship between acoustic feature and speech waveform sample points, which significantly improves the fidelity of the synthesized speech. There are also vocoders based on autoregressive structures for speech waveform sample point prediction, such as WaveNet \cite{ref_article14}, and non-autoregressive structures with high fidelity and fast generation speed, such as Hifi-GAN \cite{ref_article12}, MelGAN \cite{ref_article13}, etc. 

However, these methods often have a large amount of model parameters and training data, the training also takes a long time, which cannot be used in some scenarios with high real-time requirements.

\begin{figure}[t]
    \centering
    \includegraphics[width=8.5cm]{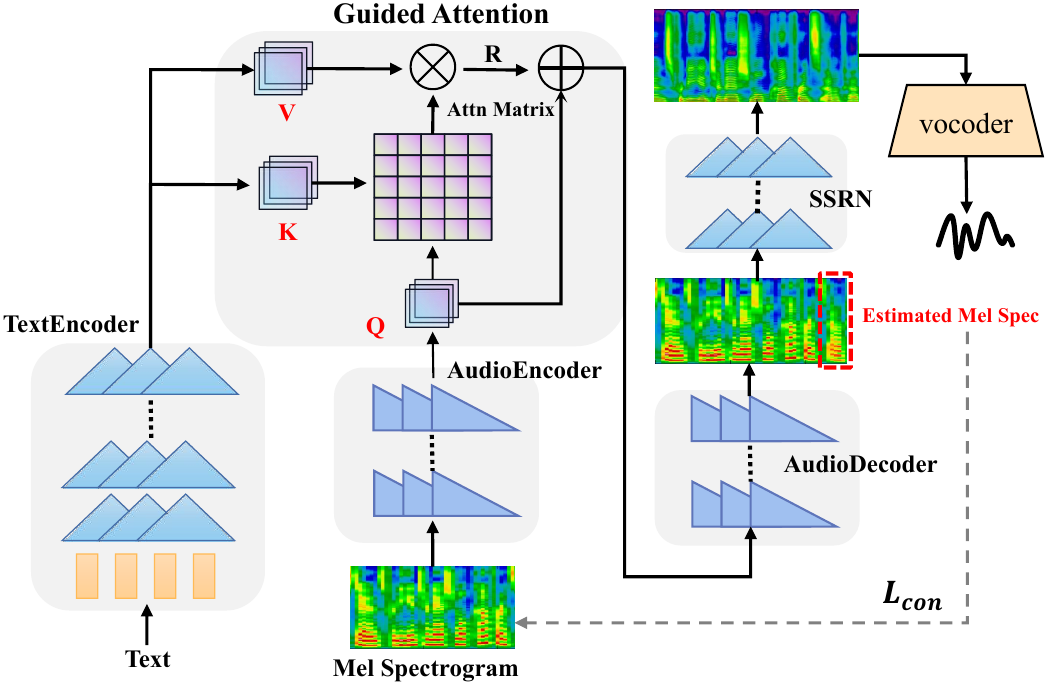}
    \caption{The architecture of our proposed EM-TTS} 
    \vspace{-0.6cm}
    \label{model_artiture}
\end{figure}

\section{Methodology}

As shown in Fig.\ref{model_artiture}, our EM-TTS model contains three modules: Text2Spectrum, SSRN, and Guided Attention, where Text2Spectrum and SSRN were used as acoustic model to synthesize the spectrum. 1) Text2Spectrum includes four components TextEncoder, AudioEncoder, Attention, and AudioDecoder, which are used to encode phonemes into coarse mel spectrogram; 2) SSRN is stacked with 1d-CNN modules to restore fine-grained spectrogram from coarse spectrogram output by Text2Spectrum; 3) Guided Attention is used to perform dimensional alignment of mel spectra and phoneme sequences.


We use a TTS system based entirely on CNN, and one of the advantages of abandoning the use of recursive unit components is that compared with the TTS system based on RNN components, the model training speed is faster while ensuring a certain degree of fidelity and naturalness, and the requirements for GPU are also lower, it is friendly to individual contestants; secondly, we use a two-stage synthesis strategy. Compared with the operation of directly synthesizing mel spectrogram or STFT spectrogram from text, we first synthesize a low-resolution coarse mel spectrogram from text, and then get Fine-Grained spectrogram from coarse spectrogram. For the high-resolution complete STFT spectrogram, the waveform file is finally obtained through a vocoder.

\subsection{Text2Spectrum Module}
We use this module to synthesize a coarse mel spectrogram from the text. The module is composed of four sub-modules: Text Encoder, Audio Encoder, Audio Decoder and Attention. Text Encoder first encodes the input text sentence 
$L=[ l_{1}, l_{2},..., l_{N}]$ into K and V, with shape $R^{d \times N}$. where d is the dimension of the encoded character. On the other hand, Audio Encoder will convert the coarse mel spectrogram of the voice with the length of $T\in R^{F\times T}$  encoded as matrix $Q\in R^{d \times T}$.
\begin{equation}
    \begin{split}
        K, V = TextEncoder(L)  \\
        Q = AudioEncoder(S_{1:F,1:T}) \\
        A = Softmax(K^{T}Q / \sqrt{d}) 
    \end{split}
\end{equation}

\normalsize
\noindent Attention matrix $A$ is used to evaluate the correlation between the $n$-th character $l_{n}$ and the $t$-th  mel spectrogram $S_{1:F,t}$. At the same time, the attention module will pay attention to the character $l_{n+1}$ in the following time and encoded to the $n$-th line of $V$. Therefore, the matrix $R$ as the subsequent mel spectrogram is defined as:
\normalsize
\begin{equation}
R = attn(Q,K,V) := V{\cdot}A
\end{equation}

\normalsize
Then, the encoded audio matrix $Q$ and matrix $R$ are spliced into $R'$, which is used as the input of audio decoder to estimate mel spectrogram.
\begin{equation}
\begin{split}
    \hat{Y}_{1:F,2:T+1} &= AudioDecoder(R') \\
    R' &= Concat[R,Q]
\end{split}
\end{equation}

\normalsize
The error of prediction results and ground truth is evaluated by loss function $\mathcal{L}_{hiera}({Y}_{1:F,2:T+1}|{s}_{1:F,2:T+1})$, which consists of $\mathcal{L}_{1}$ $Loss$ and binary divergence function $\mathcal{L}_{spec}$:

\begin{footnotesize}
\begin{equation}
    \begin{aligned}
        \mathcal{L}_{spec}(\mathbf{Y}|\mathcal{S}) &= \mathbf{E}_{ft}[-\mathcal{S}_{ft}log(\frac{\mathbf{Y}_{ft}}{{S}_{ft}}) - (1-{S}_{ft})log(\frac{1-{Y}_{ft}}{1-{S}_{ft}})] \\
        &=\mathbf{E}_{ft}[-\mathcal{S}_{ft}\hat{Y}_{ft} + log(1+e^{\hat{Y}_{ft}})]        
    \end{aligned}
\end{equation}
\end{footnotesize}
\begin{footnotesize}
\begin{equation}
\mathcal{L}_{hiera} = \mathcal{L}_{spec}(\mathbf{Y}|\mathcal{S}) + \mathbf{E}[| \mathbf{Y}_{ft} - \mathcal{S}_{ft}|]
\label{hiera}
\end{equation}
\end{footnotesize}

TextEncoder consists of a character embedding and several 1-D no-causal convolution layers. AudioEncoder and AudioDecoder are composed of 1-D causal convolution layers. These convolutions should be causal, because the output of AudioDecoder is fed back to the input of AudioEncpder in the synthesis phase.

\subsection{Spectrogram Super-Resolution Module}
In the second stage of synthesis, we use SSRN to further synthesize the complete spectrum from the coarse mel spectrogram. For frequency up sampling, we can increase the number of channels in the 1-D convolution layer, and for upsampling on the time axis, we can increase the sequence length from $T$ to four times the original length by twice deconvolution.

Since we do not consider online data processing, all convolutions in SSRN are non causal. The loss function used by this module is the same as that of Text2Spectrum in Eq.\ref{hiera}.

\subsection{Guided Attention}
Due to the correspondence between the order of text characters and audio clips, the attention module in Text-to-Speech needs to pay extra attention to the word alignment between different languages, in addition to the fact that the reading step of text characters is often performed linearly in time by default.

We use guided attention loss, which can make the attention matrix distributed near the diagonal, and set a penalty term if the attention matrix is distributed far from the diagonal, which means that characters are loaded in random order and not linearly.

\begin{equation}
\mathcal{L}_{attn}(A) = \mathbf{E}_{attn}[\mathcal{A}_{attn}\mathcal{W}_{attn}]
\end{equation}

Where $\mathcal {W}_{attn}$ in the above equation is $1-exp[-(n/N-t/T)]^{2}/2g^{2}$, where $g$ = 0.2, $\mathcal{L}_{attn}(A)$ as an auxiliary loss function, is updated iteratively with the main loss function $\mathcal{L}_{hiera}$. In our experiments, if we add bootstrap attention loss as an auxiliary loss function to the objective, it reduces the number of iterations required for training and indirectly reduces the time consumed to train the Text-to-Speech model.

During the synthesis phase, the attention matrix $A$ sometimes fails to focus on the correct characters, typically skipping a few letters and repeating the same word twice or more. To make the system more robust, we heuristically modify the attention matrix to be distributed close to the diagonal by some simple rules. This approach sometimes alleviates such problems.

\subsection{Data Augmentation}
To address the problem of low resources, we designed a series of data enhancement methods to expand the dataset. 
\subsubsection{\textbf{Noise Suppression (DN)}} 
\begin{figure}
    \centering
    \includegraphics[width=5.5cm]{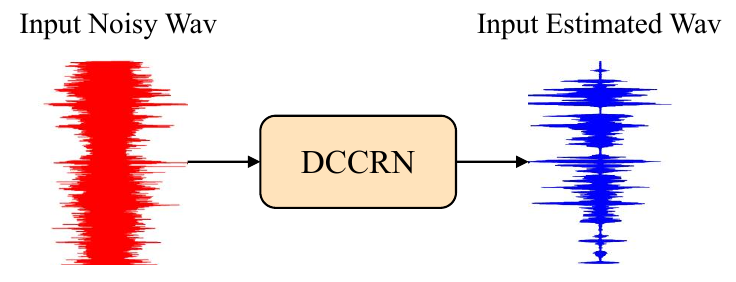}
    \vspace{-0.2cm}
    \caption{Noise Suppression using DCCRN model} 
    \vspace{-0.5cm}
    \label{dccrn}
\end{figure}

As shown in Fig.\ref{dccrn}, we use the DCCRN \cite{ref_article14} model to perform noise reduction processing on the data set, which reduces noise interference and expands the data set.

\subsubsection{\textbf{SpecArgument (SA)}} 
Facing the current situation of low resources of Mongolian language data, we consider using data augmentation to supplement the training samples and improve the model robustness, and the common methods are adding natural noise or artificial noise, volume enhancement, speed enhancement, etc. The advantage of noise augmentation is that it can make the model more robust and applicable to more scenarios, but the disadvantage is that it requires a large amount of noisy data, and insufficient data will affect the generalization ability.

In this paper, we use SpecAugment, which simply performs the three operations of Time Warping, Frequency Masking, and Time Masking on the speech spectrogram.
\vspace{-0.5cm}
\begin{figure}[htbp]
\centering
\subfigure[Mel Spectrogram.]{
\begin{minipage}[t]{0.5\linewidth}
\centering
\includegraphics[width=4.2cm]{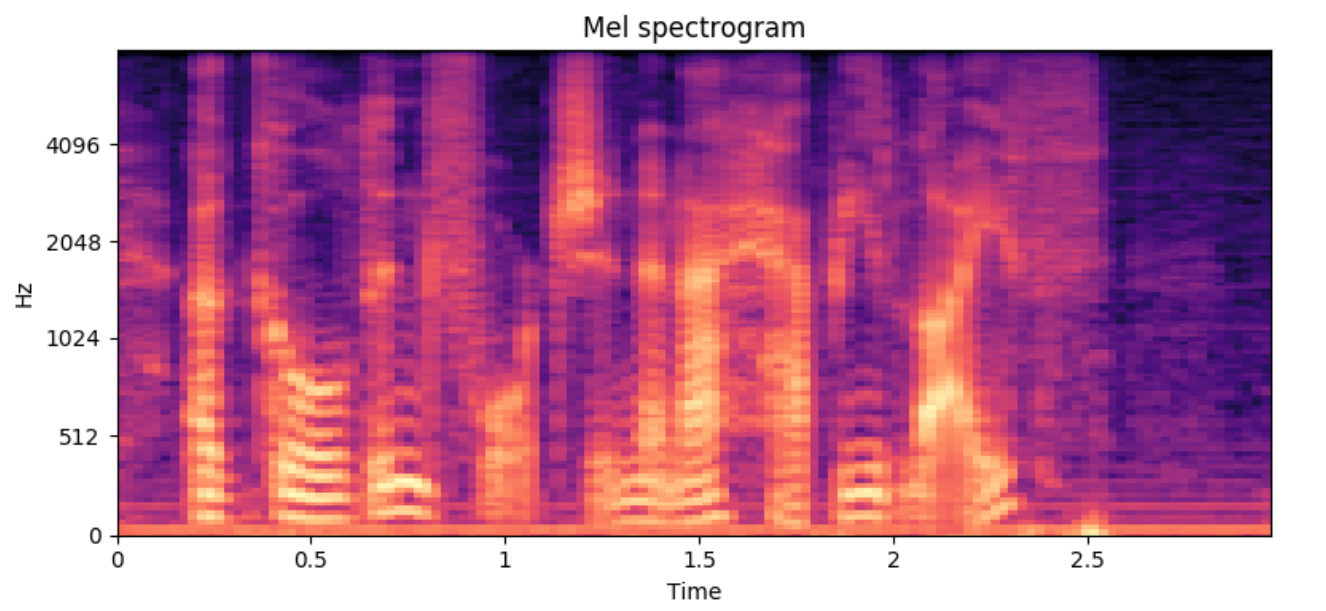}
\end{minipage}%
}%
\subfigure[Mel Spectrogram Masked.]{
\begin{minipage}[t]{0.5\linewidth}
\centering
\includegraphics[width=4.2cm]{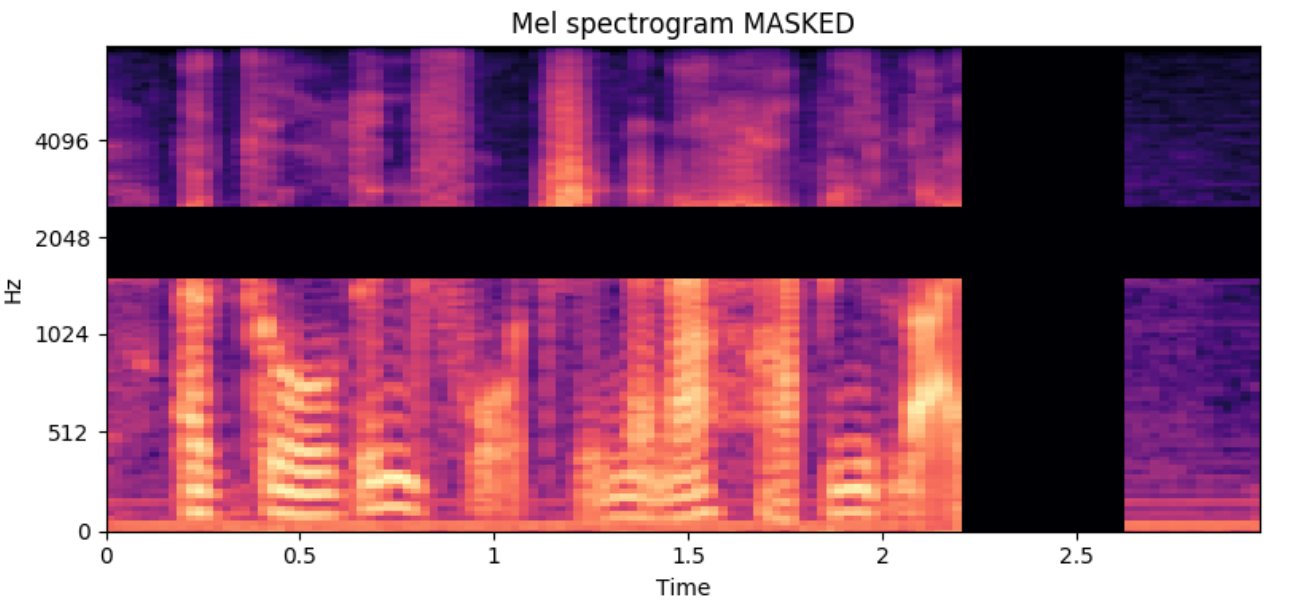}
\end{minipage}%
}%
\centering
\vspace{-0.2cm}
\caption{Mel Spectrogram by SpecArgument}
\end{figure}
\vspace{-0.2cm}

\begin{itemize}
\item[$\bullet$] \textbf{Time Warping}: A data point is randomly selected on the time axis of the speech spectrogram and randomly distorted to the left or right at a certain distance, with the distance parameter randomly chosen from a uniform distribution of time distortion parameters.
\end{itemize}

\begin{itemize}
\item[$\bullet$] \textbf{Frequency Masking}: The part of the frequency channel domain $[f_{0}, f_{0}+f)$ is masked in its entirety, where $f$ is chosen from a uniform distribution of the parameters from point 0 to the frequency mask, $f_{0}$ is chosen from $(0, v-f)$, and $v$ is the number of channels in the frequency dimension.
\end{itemize}

\begin{itemize}
\item[$\bullet$] \textbf{Time Masking}: The spectrogram data mask of $t$ consecutive time steps $[t_{0}, t_{0}+t]$ interval, where $t_{0}$ is chosen from $[0,T-t]$ interval, $T$ is the length of audio duration obtained by framing means, and likewise t is taken from 0 to the uniform distribution of time mask parameters. 
\end{itemize}

The experimental results show that the method does improve the training speed, which is due to the fact that no further data conversion is required between the waveform data to the spectrogram data, and it is a direct enhancement of the spectrogram data.

\subsubsection{\textbf{Spectrogram-Resize (SR)}} 

In addition to SpecArgument, Spectrogram-Resize data augmentation \cite{ref_article15} is used to supplement data by compressed or stretched source spectrograms to alleviate the low-resource problem of insufficient training corpus. This method is simple and easier to implement, and does not require complex signal processing knowledge.

\vspace{-0.2cm}
\begin{figure}[htbp]
  \centering
  \begin{minipage}[t]{0.48\textwidth}
    \centering
    \includegraphics[width=\textwidth]{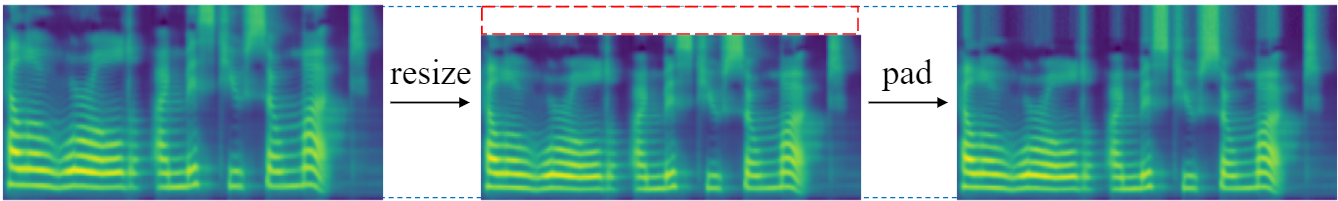}
    \caption{Resize ratio \textless 1}
    \label{resize1}
  \end{minipage}
  \begin{minipage}[t]{0.48\textwidth}
    \centering
    \includegraphics[width=\textwidth]{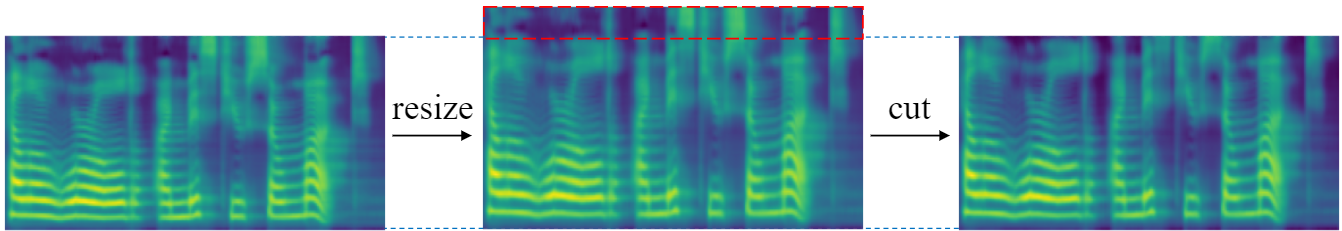}
    \caption{Resize ratio \textgreater 1}
    \vspace{-0.2cm}
    \label{resize2}
  \end{minipage}
\end{figure}

Spectrogram-Resize data augmentation operations are as follows: extract Mel spectrogram from audio waveforms, and perform vertical frequency bin axis resize or horizontal time axis resize operation on the spectrogram. Using the frequency bin axis as an example, as shown in Fig.\ref{resize1} and Fig.\ref{resize2}, the melspectrogram is first adjusted to a specific proportion by using bilinear interpolation, and then the melspectrogram is padded or cut to an original shape. If the ratio is less than 1, it produces audio with a lower pitch and a closer formant distance, and vice versa. We train the enhanced audio, and the model can better learn the content information of the speaker in the audio. In addition to the operations on the vertical frequency bin axis, the resize operation can also be performed on the horizontal time axis.

We use SpecAugment to solve the problem of overfitting and model generalization due to the lack of sufficient training data. However, it should be noted that the introduction of data augmentation turns the overfitting problem into an underfitting problem, and although the validation effect on the training set decreases, the effect on the test set does improve, which proves the merit of SpecAugment in improving the generalization of the model.

\section{Experiments and Results}

The NCMMSC2022-MTTSC \cite{ref_article0} dataset was recorded by a professional female announcer whose native language is Mongolian. The dataset contains a total of 1000 pieces, about 2 hours of audio. We use data augmentation to expand the data set to 6k pieces, about 12 hours of audio. we divide the training set and validation set according to the ratio of 8:2. The entire recording process was recorded in a standard recording studio at Inner Mongolia University using Adobe Audition software. The announcer followed our text script and read aloud sentence by sentence. In addition, a volunteer supervises the recording process and asks the announcer to re-record if there are any murmurs or unreasonable pauses during the recording process. 

\subsection{Experimental Setup}

We train on the officially provided training set and use the validation set to evaluate the model performance and select the model with the best results for inference testing. The experimental device is a 1080Ti with 12G memory. we train Text2Spectrum and SSRN for about 15h and 30h respectively. We used random search and got the best choice of hyperparameters, as shown in Table \ref{tab1}.

\begin{table}[thbp]\centering
\vspace{-0.1cm}
\caption{Parameters Set.}\label{tab1}
\vspace{-0.2cm}
\begin{tabular}{c|c}
\toprule
\cmidrule(r){1-2}
\textbf{Parameter}    & \textbf{Value}                   \\ 
\midrule
Text2Spectrum learning rate        & 0.005                      \\ 
\midrule
SSRN learning rate        & 0.0005                      \\ 
\midrule
Sampling rate         & 22050 Hz                         \\ 
\midrule
Adam $\alpha$, $\beta_{1}$, $\beta_{2}$, $\xi$   & 2e-4, 0.5, 0.9, 10e-6       \\ 
\midrule
Dimension e, d and c  & 128, 256, 512                    \\ 
\midrule
STFT spectrogram size & 80 × T (T depends on audio clip) \\ 
\midrule
Mel spectrogram size  & 80 × T (T depends on audio clip) \\ 
\bottomrule
\end{tabular}
\vspace{-0.1cm}
\end{table}

For simplicity, we train Text2Spectrum and SSRN independently and asynchronously using different GPUs. all model parameters are initialized using a Gaussian initializer. Both modules are trained by the ADAM optimizer. When training the SSRN, we randomly extract short sequences of T = 64 for each iteration to save memory usage. To reduce disk access, we reduce the frequency of creating parameters and saving the model to only once per 2K iterations.

\subsection{Results}
\begin{figure}[thbp]
    \centering
    \includegraphics[width=7.5cm]{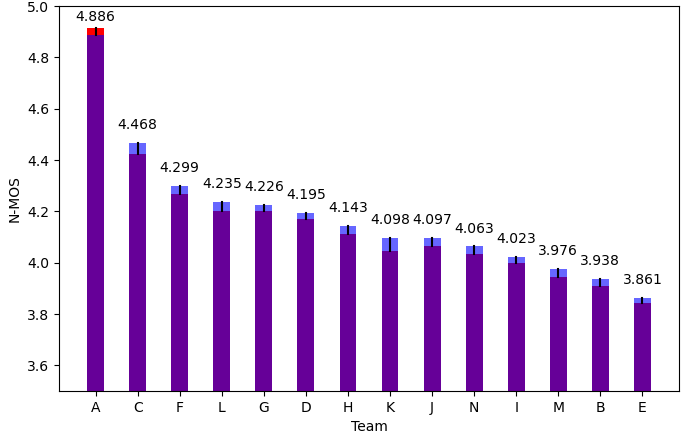}
    \includegraphics[width=7.5cm]{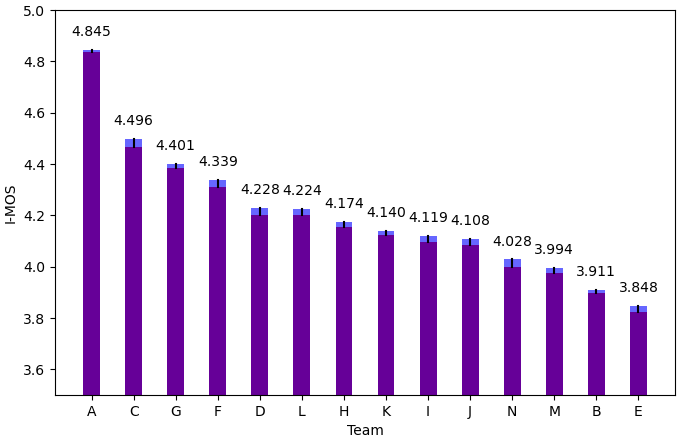}
    \vspace{-0.2cm}
    \caption{N-MOS and I-MOS Scores} 
    \vspace{-0.65cm}
    \label{N-MOS}
\end{figure}

The system evaluation was conducted with 20 Mongolian-speaking volunteers to assess speech naturalness and speaker similarity of speech synthesized from different models. The speech was presented to volunteers in random order, and volunteers were asked to rate the speech on a scale of 1 to 5, with 1-point intervals. The evaluation metrics are defined as Naturalness Mean Opinion Score (N-MOS) and Intelligibility Mean Opinion Score (I-MOS). The test results of our submitted inference audio are as given in Fig.\ref{N-MOS}.

Our EM-TTS solution ranks 12th among all participating teams in terms of N-MOS and I-MOS metrics (TeamNumber B), where A is the MOS score of bonafide audio (Ground-Truth). It can be seen that our proposed system has N-MOS score of 3.938 and I-MOS score of 3.911, which has the great advantage of significantly reducing the training time and is not very demanding on the experimental equipment, while ensuring an acceptable degree of naturalness and intelligibility.


We conducted several groups of experiments on Mongolian datasets provided by the organizer. Tacotron (based on RNN), Tacotron2 (based on RNN), Choi et al.\cite{Choi2019MultispeakerEA} (based on CNN), and Fastspeech2 (Based on Transformer) were used as acoustic models, and HiFiGAN v1 generators were used as vocoders for training on Mongolian datasets. Finally, the trained model is used for joint reasoning, and the experimental results of several groups of schemes are evaluated. 

\begin{table}[bp]
  \vspace{-0.3cm}
  \caption{Comparison of different methods in Evaluation.}
  \label{tab2}
  \centering
  \begin{tabular}{c | c | c | c}
    \toprule
    \cmidrule(r){1-4}
    \textbf{Methods}  &\textbf{MOS}$\uparrow$  &\textbf{SMOS}$\uparrow$  &\textbf{MCD}$\downarrow$  \\ 
    \midrule
     Tacotron + Griffin-Lim    & $3.07 \pm 0.12$    &$3.01 \pm 0.13$    & 6.86     \\ 
    \midrule
     Tacotron2 + HifiGAN       & $4.01 \pm 0.08$    &$3.89 \pm 0.12$    & 5.12      \\ 
    \midrule
     Choi et al.\cite{Choi2019MultispeakerEA} + HifiGAN   & $3.88 \pm 0.11$  &$3.52 \pm 0.08$  & 6.04  \\ 
    \midrule
     Fastspeech2 + HifiGAN     & $3.93 \pm 0.09$    &$3.73 \pm 0.11$    & 5.80      \\ 
    \midrule
    \textbf{Ours} &\textbf{$3.92 \pm 0.08$ }  &\textbf{$3.77 \pm 0.09$}  & 5.47    \\ 
    \bottomrule
  \end{tabular}
  \vspace{-0.5cm}
\end{table}

As shown in Table \ref{tab2}, subjective tests are used to evaluate the speech naturalness (MOS) and speaker similarity (SMOS) of converted speeches generated from different models. We also calculate mel-cepstrum distortion (MCD) \cite{mcd} as objective tests for converted audio and target audio to evaluate our TTS model. 
The results show that we achieved a naturalness that is close to FastSpeech2+HifiGAN (FS2+HF) and Tacotron2+HifiGAN (TA2+HF), at the same time, the speaker similarity exceeds FS2+HF, Tacotron+Griffin-Lim (TA+GLA), and Choi et al.\cite{Choi2019MultispeakerEA}. This demonstrates that our model can generate speech with good naturalness and high similarity.
The MCD of our method is lower than FastSpeech2, TA+GLA, and CNN-based TTS \cite{Choi2019MultispeakerEA}, which proves that the audio synthesized by our method has good quality to a certain extent.
It shows that TTS with a fully CNN structure can completely approximate TTS with RNN or Transformer structure in terms of performance, while reducing the economic cost of training.


\begin{table}[thbp]
\centering
\vspace{-0.1cm}
\caption{Comparison of Times, WER, PCC, and Param.}\label{tab3}
\begin{tabular}{c|c|c|c|c}
\toprule
\cmidrule(r){1-5}
\textbf{}    &\textbf{Choi et al.}   &\textbf{TA2+HF}   &\textbf{FS2+HF}  &\textbf{Ours} \\ 
\midrule
\textbf{WER(\%)}$\downarrow$      & $26.19$  & $24.87$  & $21.60$   &$22.94$             \\ 
\midrule
\textbf{log F0-PCC}$\uparrow$     & $0.665$  & $0.691$  & $0.787$   &$0.743$              \\ 
\midrule
\textbf{Times (h)}$\downarrow$    & $\approx35$  & $\approx56$  & $\approx25$  & $\approx15$          \\ 
\midrule
\textbf{Param.Count (M)}          & $215.4$   & $376.2$  & $387.6$   & $156.5$               \\ 
\bottomrule
\end{tabular}
\vspace{-0.2cm}
\end{table}

As shown in Table \ref{tab3}, we calculate the training times, model parameters, Word Error Rate (WER), and the Pearson correlation coefficient (PCC) \cite{pcc} between F0 of the target and the synthesized speech. 
Although the audio quality and fidelity of our model are not as good as Tacotron2 and FastSpeech2, the model parameters and  training speed of our model completely overwhelms other solutions. In addition, EM-TTS inference results are basically the same as fastspeech2 in terms of sound quality and fidelity, but the training time is improved by 40\%. Our model significantly reduces the amount of model parameters and reduces the GPU memory requirements for model deployment, making it possible to use the model on mobile terminals.

What's more, our method is better than Tacotron2 on WER, which shows that our model can retain relatively complete content information and achieve content consistency when performing speech synthesis. At the same time, according to PCC, we can achieve a good fitting of F0 and achieve the effect of synthesizing the speaker's pitch.

\subsection{Ablation Study}
We performed ablation experiments on the three data augmentation techniques employed in our solution in Table \ref{tab4}. 
\begin{table}[htbp]\centering
\vspace{-0.3cm}
\caption{Comparison of three data augmentation.}\label{tab4}
\begin{tabular}{c|c|c|c}
\toprule
\cmidrule(r){1-4}
\textbf{Methods}  &\textbf{MOS}$\uparrow$  &\textbf{SMOS}$\uparrow$   &\textbf{MCD}$\downarrow$           \\ 
\midrule
w/o SA    & $3.893 \pm 0.08$       & $3.681 \pm 0.11$     & $6.77$      \\ 
\midrule 
w/o SR    & $3.882 \pm 0.06$       & $3.636 \pm 0.09$     & $6.26$       \\ 
\midrule
w/o DN    & $3.923 \pm 0.09$       & $3.711 \pm 0.12$     & $5.91$       \\ 
\bottomrule
\end{tabular}
\end{table}
\vspace{-0.2cm}

The experimental results indicate that training on the augmented dataset leads to an improvement in the audio quality, fidelity, and similarity of the speech generated by EM-TTS inference, regardless of whether to use a single data augmentation technique, or a combination of data augmentation. This demonstrates the effectiveness of our three proposed three data augmentations. While expanding the training data, it enhances the robustness of the model and improves the quality of speech synthesis of the TTS model.

\subsection{Discussion}
We uses the TTS model based on the full CNN and the technology for fast training of Guided attention module. Compared with the mainstream solution of using Tacotron2 based on RNN or FastSpeech2 based on transformer as the acoustic model, the training accuracy is definitely less accurate. However, this solution using full convolution has achieved results close to the former under subjective and objective evaluations. Our EM-TTS surpassed Tacotron and FastSpeech2 in the MCD metrics, and defeated the TTS model based on RNN in WER and PCC metrics, which significantly reduces model parameters and training time consumption simultaneously.

\section{Conclusion}
In this paper, we present a lightweight TTS model that significantly reduces training time and model parameters while maintaining a certain synthesis quality. Our proposed two-stage training modules Text2Spectrum and SSRN implement the conversion of phonemes to coarse mel spectrogram and coarse spectrogram to fine-grained spectrogram, respectively, enabling high-efficiency speech synthesis. Experimental results show that while the audio quality is far from perfect, our model takes significantly less time to train and reduces a large number of model parameters which does not require a lot of computational resources and alleviate these economic costs of training. 

\section{Acknowledgement}
Thanks for the advice from Xulong Zhang from Ping An Technology (Shenzhen) Co., Ltd. (zhangxulong@ieee.org). His valuable guidance and support greatly facilitated the completion of this article.

\bibliographystyle{IEEEtran.bst}
\bibliography{EM-TTS.bib}

\begin{thebibliography}{10}
\providecommand{\url}[1]{#1}
\csname url@samestyle\endcsname
\providecommand{\newblock}{\relax}
\providecommand{\bibinfo}[2]{#2}
\providecommand{\BIBentrySTDinterwordspacing}{\spaceskip=0pt\relax}
\providecommand{\BIBentryALTinterwordstretchfactor}{4}
\providecommand{\BIBentryALTinterwordspacing}{\spaceskip=\fontdimen2\font plus
\BIBentryALTinterwordstretchfactor\fontdimen3\font minus \fontdimen4\font\relax}
\providecommand{\BIBforeignlanguage}[2]{{%
\expandafter\ifx\csname l@#1\endcsname\relax
\typeout{** WARNING: IEEEtran.bst: No hyphenation pattern has been}%
\typeout{** loaded for the language `#1'. Using the pattern for}%
\typeout{** the default language instead.}%
\else
\language=\csname l@#1\endcsname
\fi
#2}}
\providecommand{\BIBdecl}{\relax}
\BIBdecl

\bibitem{ref_article1}
L.~Rui, B.~Feilong, G.~Guanglai, and W.~Yonghe, ``Mongolian text-to-speech system based on deep neural network,'' \emph{National conference on Man-Machine Speech Communication, {NCMMSC}}, pp. 99--108, 2018.

\bibitem{ref_article5}
Y.~Fan, Y.~Qian, F.~Xie, and F.~K. Soong, ``{TTS} synthesis with bidirectional {LSTM} based recurrent neural networks,'' in \emph{15th Annual Conference of the International Speech Communication Association {Interspeech}}, 2014, pp. 1964--1968.

\bibitem{ref_article6}
H.~Zen and H.~Sak, ``Unidirectional long short-term memory recurrent neural network with recurrent output layer for low-latency speech synthesis,'' in \emph{{IEEE} International Conference on Acoustics, Speech and Signal Processing, {ICASSP}}, 2015, pp. 4470--4474.

\bibitem{ref_article7}
Y.~Wang, R.~J. Skerry{-}Ryan, D.~Stanton, Y.~Wu, R.~J. Weiss, N.~Jaitly, Z.~Yang, Y.~Xiao, Z.~Chen, S.~Bengio, Q.~V. Le, Y.~Agiomyrgiannakis, R.~Clark, and R.~A. Saurous, ``Tacotron: Towards end-to-end speech synthesis,'' in \emph{18th Annual Conference of the International Speech Communication Association, {Interspeech}}, 2017, pp. 4006--4010.

\bibitem{fastspeech1}
Y.~Ren, Y.~Ruan, X.~Tan, T.~Qin, S.~Zhao, Z.~Zhao, and T.~Liu, ``Fastspeech: Fast, robust and controllable text to speech,'' in \emph{Annual Conference on Neural Information Processing Systems, {NeurIPS}}, 2019, pp. 3165--3174.

\bibitem{ref_article10}
Y.~Ren, C.~Hu, X.~Tan, T.~Qin, S.~Zhao, Z.~Zhao, and T.~Liu, ``Fastspeech 2: Fast and high-quality end-to-end text to speech,'' in \emph{9th International Conference on Learning Representations, {ICLR}}, 2021.

\bibitem{ref_article11}
J.~Kong, J.~Kim, and J.~Bae, ``Hifi-gan: Generative adversarial networks for efficient and high fidelity speech synthesis,'' in \emph{Annual Conference on Neural Information Processing Systems, {NeurIPS}}, 2020.

\bibitem{parallelGAN}
R.~Yamamoto, E.~Song, and J.~Kim, ``Parallel wavegan: {A} fast waveform generation model based on generative adversarial networks with multi-resolution spectrogram,'' in \emph{{IEEE} International Conference on Acoustics, Speech and Signal Processing, {ICASSP}}, 2020, pp. 6199--6203.

\bibitem{vits}
J.~Kim, J.~Kong, and J.~Son, ``Conditional variational autoencoder with adversarial learning for end-to-end text-to-speech,'' in \emph{Proceedings of the 38th International Conference on Machine Learning, {ICML}}, vol. 139, 2021, pp. 5530--5540.

\bibitem{yourtts}
E.~Casanova, J.~Weber, C.~D. Shulby, A.~C. J{\'{u}}nior, E.~G{\"{o}}lge, and M.~A. Ponti, ``Yourtts: Towards zero-shot multi-speaker {TTS} and zero-shot voice conversion for everyone,'' in \emph{International Conference on Machine Learning, {ICML}}, vol. 162, 2022, pp. 2709--2720.

\bibitem{naturalspeech}
X.~Tan, J.~Chen, H.~Liu, J.~Cong, C.~Zhang, Y.~Liu, X.~Wang, Y.~Leng, Y.~Yi, L.~He, F.~K. Soong, T.~Qin, S.~Zhao, and T.~Liu, ``Naturalspeech: End-to-end text to speech synthesis with human-level quality,'' \emph{CoRR}, vol. abs/2205.04421, 2022.

\bibitem{valle}
C.~Wang, S.~Chen, Y.~Wu, Z.~Zhang, L.~Zhou, S.~Liu, Z.~Chen, Y.~Liu, H.~Wang, J.~Li, L.~He, S.~Zhao, and F.~Wei, ``Neural codec language models are zero-shot text to speech synthesizers,'' \emph{CoRR}, vol. abs/2301.02111, 2023.

\bibitem{speartts}
E.~Kharitonov, D.~Vincent, Z.~Borsos, R.~Marinier, S.~Girgin, O.~Pietquin, M.~Sharifi, M.~Tagliasacchi, and N.~Zeghidour, ``Speak, read and prompt: High-fidelity text-to-speech with minimal supervision,'' \emph{CoRR}, vol. abs/2302.03540, 2023.

\bibitem{gradtts}
V.~Popov, I.~Vovk, V.~Gogoryan, T.~Sadekova, and M.~A. Kudinov, ``Grad-tts: {A} diffusion probabilistic model for text-to-speech,'' in \emph{Proceedings of the 38th International Conference on Machine Learning, {ICML}}, vol. 139, 2021, pp. 8599--8608.

\bibitem{difftts}
M.~Jeong, H.~Kim, S.~J. Cheon, B.~J. Choi, and N.~S. Kim, ``Diff-tts: {A} denoising diffusion model for text-to-speech,'' in \emph{22nd Annual Conference of the International Speech Communication Association, {Interspeech}}, 2021, pp. 3605--3609.

\bibitem{guidetts}
H.~Kim, S.~Kim, and S.~Yoon, ``Guided-tts: {A} diffusion model for text-to-speech via classifier guidance,'' in \emph{International Conference on Machine Learning, {ICML}}, vol. 162, 2022, pp. 11\,119--11\,133.

\bibitem{PriorGrad}
S.~Lee, H.~Kim, C.~Shin, X.~Tan, C.~Liu, Q.~Meng, T.~Qin, W.~Chen, S.~Yoon, and T.~Liu, ``Priorgrad: Improving conditional denoising diffusion models with data-dependent adaptive prior,'' in \emph{The Tenth International Conference on Learning Representations, {ICLR}}, 2022.

\bibitem{zhang2022semi}
X.~Zhang, J.~Wang, N.~Cheng, and J.~Xiao, ``Semi-supervised learning based on reference model for low-resource tts,'' in \emph{2022 18th International Conference on Mobility, Sensing and Networking, {MSN}}.\hskip 1em plus 0.5em minus 0.4em\relax IEEE, 2022, pp. 966--971.

\bibitem{zhang2022tdass}
{Zhang, Xulong and Wang, Jianzong and Cheng, Ning and Xiao, Jing}, ``Tdass: Target domain adaptation speech synthesis framework for multi-speaker low-resource tts,'' in \emph{2022 International Joint Conference on Neural Networks, {IJCNN}}.\hskip 1em plus 0.5em minus 0.4em\relax IEEE, 2022, pp. 1--7.

\bibitem{ref_article14}
A.~van~den Oord, S.~Dieleman, H.~Zen, K.~Simonyan, O.~Vinyals, A.~Graves, N.~Kalchbrenner, A.~W. Senior, and K.~Kavukcuoglu, ``Wavenet: {A} generative model for raw audio,'' in \emph{The 9th {ISCA} Speech Synthesis Workshop, {SSW}}, 2016, p. 125.

\bibitem{ref_article12}
K.~Kumar, R.~Kumar, T.~de~Boissiere, L.~Gestin, W.~Z. Teoh, J.~Sotelo, A.~de~Br{\'{e}}bisson, Y.~Bengio, and A.~C. Courville, ``Melgan: Generative adversarial networks for conditional waveform synthesis,'' in \emph{Annual Conference on Neural Information Processing Systems, {NeurIPS}}, 2019, pp. 14\,881--14\,892.

\bibitem{ref_article13}
J.~Shen, R.~Pang, R.~J. Weiss, M.~Schuster, N.~Jaitly, Z.~Yang, Z.~Chen, Y.~Zhang, Y.~Wang, R.~Ryan, R.~A. Saurous, Y.~Agiomyrgiannakis, and Y.~Wu, ``Natural {TTS} synthesis by conditioning wavenet on {MEL} spectrogram predictions,'' in \emph{{IEEE} International Conference on Acoustics, Speech and Signal Processing, {ICASSP}}, 2018, pp. 4779--4783.

\bibitem{ref_article15}
N.~Kalchbrenner, E.~Elsen, K.~Simonyan, S.~Noury, N.~Casagrande, E.~Lockhart, F.~Stimberg, A.~van~den Oord, S.~Dieleman, and K.~Kavukcuoglu, ``Efficient neural audio synthesis,'' in \emph{Proceedings of the 35th International Conference on Machine Learning, {ICML}}, vol.~80, 2018, pp. 2415--2424.

\bibitem{ref_article0}
``Mongolian text-to-speech challenge under low-resource scenario.'' in \emph{http://mglip.com/challenge/NCMMSC2022-MTTSC}.

\bibitem{Choi2019MultispeakerEA}
H.~Choi, S.~Park, J.~Park, and M.~Hahn, ``Multi-speaker emotional acoustic modeling for cnn-based speech synthesis,'' \emph{{IEEE} International Conference on Acoustics, Speech and Signal Processing, {ICASSP}}, pp. 6950--6954, 2019.

\bibitem{mcd}
T.~Toda, A.~W. Black, and K.~Tokuda, ``Voice conversion based on maximum-likelihood estimation of spectral parameter trajectory,'' \emph{{IEEE} Trans. Speech Audio Process, {TASLP}}, vol.~15, no.~8, pp. 2222--2235, 2007.

\bibitem{pcc}
J.~Benesty, J.~Chen, and Y.~Huang, ``On the importance of the pearson correlation coefficient in noise reduction,'' \emph{{IEEE} Trans. Speech Audio Process, {TASLP}}, vol.~16, no.~4, pp. 757--765, 2008.

\end{thebibliography}

\end{document}